\documentclass[10.5pt]{iopart}

\usepackage{graphicx}
\usepackage{bm}
\usepackage{appendix}
\usepackage{color}
\usepackage{braket}

\usepackage{multicol}
\usepackage{mathrsfs}
\usepackage{cite}
\usepackage{amsmath,amssymb}

\begin{document}

\title[]{
Magnetoresistance in the Extreme Quantum Limit: Field-Induced Crossover to the Unitarity Limit
}

\author{Shuto Tago$^{1, 2}$}
\author{Akiyoshi Yamada$^{1, 2}$}
\author{Yuki Fuseya$^{1, 2}$}

\address{$^1$Department of Physics, Kobe University, Kobe 657-8501, Japan\\
$^2$Department of Engineering Science, University of Electro-Communications, Chofu, Tokyo 182-8585, Japan\\}

\ead{shuto.tago@stu.kobe-u.ac.jp}

\vspace{10pt}
\begin{indented}
\item[]\date{\today}
\end{indented}

\begin{abstract}
    We theoretically investigate magnetoresistance (MR) in the extreme quantum limit (EQL), where the kinetic energy becomes significantly smaller than the cyclotron energy, using the Kubo formula with Green's functions and the $T$-matrix approximation. 
    We uncover a magnetic-field-induced crossover in the scattering rate: $1/\tau \propto B^2$ in the Born regime and $1/\tau \propto B^{-2}$ in the unitarity limit.
    This crossover gives rise to distinct MR behaviors in the EQL, characterized by linear transverse MR ($\rho_{xx} \propto B$) and negative longitudinal MR ($\rho_{zz} \propto B^{-2}$). 
    This dichotomy implies insulating behavior when the magnetic field is perpendicular to the current, and metallic behavior when it is parallel. In the unitarity limit, we further derive a universal relation that enables direct experimental determination of the impurity density from $\rho_{xx}$ and $\rho_{xy}$. Our results establish a quantum--classical correspondence that remains valid even in the EQL, provided that the field dependences of the scattering rate and quantum corrections are properly incorporated.
\end{abstract}

\ioptwocol

\section{Introduction}
\begin{figure}[tbp]
  \centering
  \includegraphics[width=8cm]{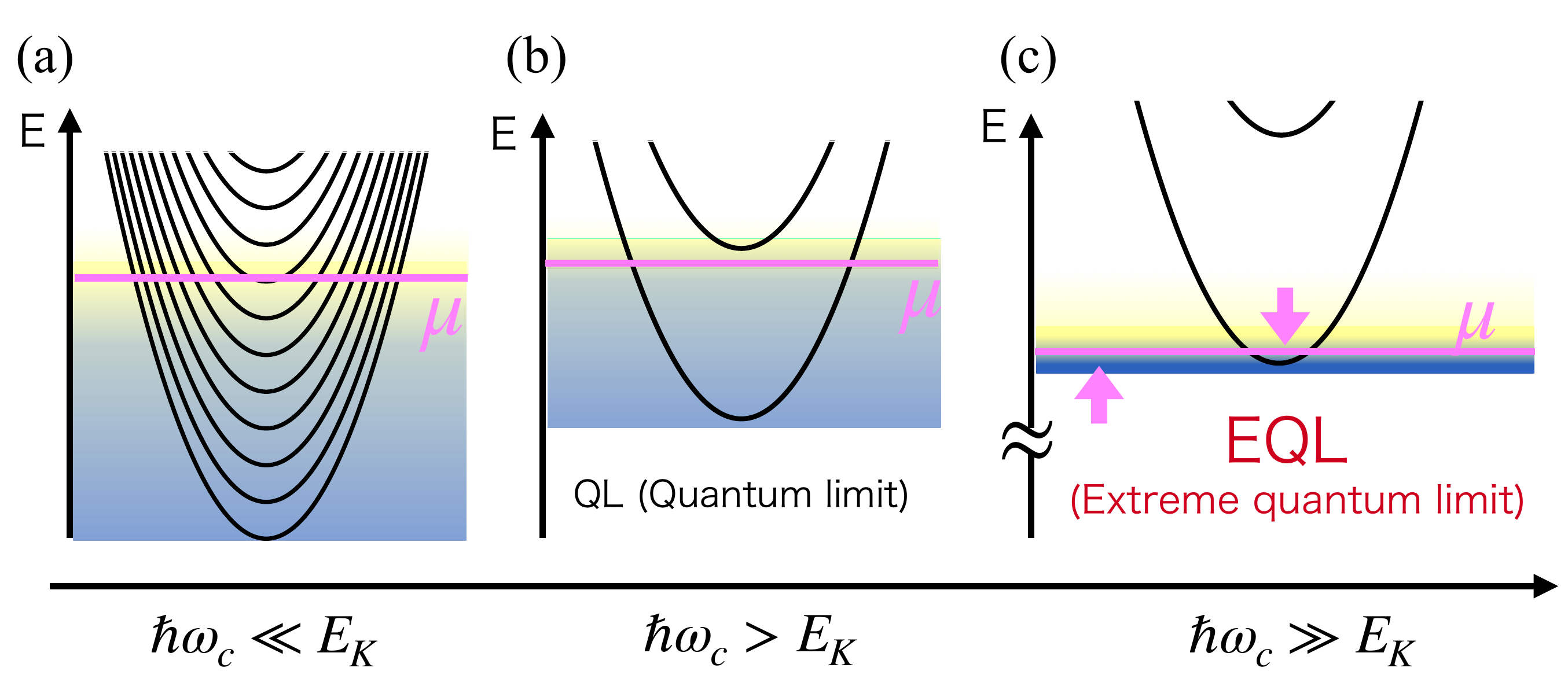}
\caption{Schematic illustrations of Landau quantization: (b) below the quantum limit (QL), (b) at the QL, and (c) in the extreme quantum limit (EQL), where the kinetic energy becomes much smaller than the cyclotron energy $\hbar \omega_c$.}\label{fig:quantum limit}
\end{figure}

Applying a magnetic field is one of the most powerful and direct methods to probe quantum effects experimentally. A quintessential example is Landau quantization, where the energy of electrons is quantized under a magnetic field, giving rise to quantum oscillations~\cite{Landau1930,Shoenberg_book}. The spacing between these Landau levels is determined by the cyclotron energy, $\hbar \omega_c = \hbar eB/m^*$, where $\hbar$ is the reduced Planck constant, $e$ ($>0$) is the elementary charge, $B$ is the magnetic field strength, and $m^*$ is the effective mass.
When the magnetic field becomes sufficiently strong, the system enters the quantum limit (QL), where the cyclotron energy $\hbar \omega_c$ exceeds the kinetic energy of electrons, $E_K$, defined as the energy difference between the chemical potential and the bottom energy of the lowest Landau level. In the QL, only the lowest Landau level remains occupied, as illustrated in Fig. \ref{fig:quantum limit}(b).

A remarkable regime emerges after the system enters the QL. As the chemical potential shifts downward to maintain a constant electron number due to the Landau level degeneracy, $E_K$ progressively decreases. When $E_K$ becomes significantly smaller than $\hbar \omega_c$, the system reaches the ``extreme quantum limit (EQL)" as depicted in Fig. \ref{fig:quantum limit}(c)~\cite{Hu2008-jk, Bhattacharya2016-yv, Kozii2019-ca, Konye_2018}. 

The nature of the EQL remains a profound mystery. In this regime, the kinetic energy approaches zero, suggesting that electrons should become nearly immobile, which would suppress electrical conductivity. 
In contrast, the reduction of $E_K$ implies a shrinking of the Fermi wave number and a corresponding increase in the Fermi wavelength, causing the electron wave function to extend over the entire system. This delocalization could, in principle, enhance conductivity.
Therefore, it is unclear whether the system in the EQL behaves as a metal or an insulator --- a fundamental and unresolved question for EQL.

In this study, we address this open problem by theoretically investigating the resistivity in the EQL. A rigorous understanding of magnetoconductivity in this regime requires a fully quantum treatment based on the Kubo formula under a strong magnetic field~\cite{Kubo_formula,Kubo_solid,Fukuyama_1969, Shiba_1971,Fuseya_2015}.
Recent progress in quantum magnetoconductivity theory has renewed the formulation of Kubo theory under a strong magnetic field and revealed a striking quantum--classical correspondence, even under strong magnetic fields where quantum oscillations are clearly visible~\cite{Yamada_2024}. Surprisingly, the classical magnetoconductivity formula remains valid even in this regime, except for a quantum correction factor that accounts for the quantum oscillations. This quantum--classical correspondence provides valuable physical insight into magnetoconductivity in EQL.
Here, we calculate the magnetoconductivity by using this reformulated Kubo formula and exploit the newly established quantum--classical correspondence to uncover the fundamental transport properties in the EQL.

\section{Theory}

\subsection{Landau levels for the effective mass approximation}

We consider a system described by the effective mass approximation. Under a magnetic field, the electron energy is quantized into Landau levels as
\begin{eqnarray} \label{eq:free_energy}
  E_{\ell}(k_z) = \left(\ell + \frac{1}{2}\right)\hbar\omega_c + \frac{\hbar^2k_z^2}{2m^\ast}, 
\end{eqnarray}
where $\ell$ is the Landau level index and $k_z$ is the wavenumber along the magnetic field, which is set parallel to the $z$-axis. Each Landau level is spin-degenerate as the Zeeman splitting is not considered in this analysis.
In the QL, the chemical potential $\mu$ satisfies $\mu < \frac{3}{2}\hbar \omega_c$, and the kinetic energy $E_K$ becomes smaller than the cyclotron energy $E_K<\hbar \omega_c$. 
The effective mass in Eq. (\ref{eq:free_energy}) corresponds to the curvature of the band edge and remains constant with respect to Fermi energy or magnetic field within the effective mass approximation or $k.p$ theory in Luttinger-Kohn representation~\cite{Luttinger1955-nd}. Even in Dirac electron systems, the non-relativistic approximation yields Eq. (\ref{eq:free_energy}) (except for the spin), which is justified when the kinetic energy is sufficiently smaller than the band gap~\cite{Wolff1964-wc}.

\subsection{Magnetoconductivity based on the Kubo formula}
The magnetoconductivity tensor$\sigma_{\mu \nu}$ is derived from the Kubo formula~\cite{Kubo_formula,Fuseya_2015,Yamada_2024}:
\begin{eqnarray}\label{eq:kubo_sigma}
\sigma_{\mu\nu} &=& {-}\lim_{\omega \to 0}  \frac{1}{{i}\omega}\left[ \Phi_{\mu\nu}(\omega) - \Phi_{\mu\nu}(0) \right].
\end{eqnarray}
Here, $\Phi_{\mu\nu}$ is the current-current correlation function, which takes the form in Matsubara representation~\cite{Matsubara_1955}:
\begin{eqnarray}
\label{eq:sigma_temp_func}
    \Phi_{\mu\nu}(i\omega_\lambda) &=& \frac{e^2}{V}\sum_{i,j}\bra{i}\pi_{\mu}\ket{j}\bra{j}\pi_{\nu}\ket{i}\mathscr{K}(i\omega_\lambda, i, j), \\
    \label{eq:sigma_green_func}
    \mathscr{K}(i \omega_\lambda, i, j) &=& {k_\mathrm{B}T}\sum_n\mathscr{G}_j(i\varepsilon_n)\mathscr{G}_i(i\varepsilon_n - i\omega_\lambda),
\end{eqnarray}
where $V$ is the system volume, $k_\mathrm{B}$ is the Boltzmann constant, and $\mathscr{G} = (i\varepsilon_n - \mathscr{H})^{-1}$ is Green's function with the Hamiltonian $\mathscr{H}$. Here, $\varepsilon_n$ and $\omega_\lambda$ are Matsubara frequencies. 
The kinematical momentum operator under a magnetic field is
$\pi_\mu = -i\hbar\partial/\partial x_\mu+ eA_\mu$, where $A_\mu$ is the vector potential, satisfying the commutation relation $[\pi_\mu, \pi_\nu] = -i\hbar e \epsilon_{\mu\nu\lambda}B_\lambda$.
Using the standard analytic continuation procedure, the Green's function in the Matsubara representation can be rewritten in terms of the retarded (advanced) Green's function $G_\ell^{\mathrm{\mathrm{R}(\mathrm{A})}}(\varepsilon, k_z) = (\varepsilon + E_\ell(k_z) \pm i{\hbar/2\tau})^{-1}$, where $\hbar/2\tau$ is the imaginary part of the self-energy and $\tau$ corresponds to the relaxation time~\cite{AGD, mahan, Fukuyama_1969,Fuseya_2015}.
Under a strong magnetic field ($\omega_c\tau \gg 1$), the total magnetoconductivity at zero temperature, including both Fermi surface and sea terms~\cite{Yamada_2024}, is given as follows:
\begin{eqnarray}
  \sigma_{xx} &=& \frac{e^2}{\pi^2\hbar\tau}\sqrt{\frac{m^*}{2}}\sum_{\ell}\left( \ell + \frac{1}{2} \right)\mathrm{Re}\left[ \frac{1}{\sqrt{K_\ell }} \right], \label{eq:sigmaxx} \\
  \sigma_{yx}&=& \frac{e^2\sqrt{2m^*}}{\pi^2\hbar^2} \sum_{\ell} \left( \mathrm{Re}\left[ \sqrt{K_\ell} \right] + \frac{\hbar}{4\tau} \mathrm{Im}\left[ \frac{1}{\sqrt{K_\ell}} \right] \right), \label{eq:sigmayx} \\
  \sigma_{zz} &=& \frac{e^2 \omega_c\tau\sqrt{2m^*}}{\pi^2\hbar^2}  \sum_{\ell} \left(\mathrm{Re}\left[ \sqrt{K_\ell}\right] + \frac{\hbar}{2\tau}\mathrm{Im}\left[ \frac{1}{\sqrt{K_\ell}} \right] \right)\label{eq:sigmazz},
\end{eqnarray}
where
\begin{eqnarray}
  K_\ell= {\mu - \left(\ell + \frac{1}{2}\right)\hbar\omega_c + i\frac{\hbar}{2\tau}} \label{eq:K_l},
\end{eqnarray}
and $\mu$ is the chemical potential.
Note that the above results are gauge independent.
The magnetoresistivity tensor can be obtained as the inverse tensor of the magnetoconductivity tensor as
\begin{eqnarray}
    \hat{\rho} = \left[\hat{\sigma} \right]^{-1}
\end{eqnarray}

\subsection{Field dependence of chemical potential}
In three-dimensional systems, the chemical potential $\mu$ remains nearly unchanged under a weak magnetic field. However, in the QL, $\mu$ changes significantly to keep the carrier number constant because the Landau level degeneracy increases linearly with the magnetic field as $N_L = eB/2\pi \hbar$. Therefore, calculating the magnetic field dependence of $\mu$ is essential for understanding properties in the QL.

The field dependence of $\mu$ is determined by the requirement that the carrier density $n $ remains constant and equal to its zero-field value $n_0 = (1/3\pi^2)\left( \sqrt{2m^\ast\mu_0}/\hbar \right)^3$, where $\mu_0$ represents the chemical potential at zero field.
At zero temperature, the carrier density is given by the standard expression $n = V^{-1}\int{_0^{\mu}} \mathscr D(\varepsilon)\mathrm{d\varepsilon}$, where $\mathscr{D}(\varepsilon)$ is the density of states and the energy dependence of $\tau$ is neglected in $\mathscr{D}(\varepsilon)$. Within the Green's function formalism, it takes the form
$
  \mathscr D(\varepsilon) = -(1/\pi)\sum_{i}\mathrm{Im} \,G^\mathrm{R}_i(\varepsilon)
$
~\cite{rickayzen_1980}.
Using this expression, the carrier density can be evaluated as
\begin{eqnarray}
  n  = \frac{3n_0\hbar\omega_c}{2\mu_0^{3/2}}\sum_\ell \mathrm{Re}\left[\sqrt{K_\ell }\right],\label{eq:carrier_green}
\end{eqnarray}
which provides the self-consistent condition to determine the field dependence of $\mu$.
Note that $K_\ell$ in Eq.~(\ref{eq:carrier_green}) implicitly depends on the scattering rate $1/\tau$, whose field dependence will be discussed in the following section.

\subsection{Field dependence of scattering rate}
\begin{figure}[tb]
  \includegraphics[width=8.2cm]{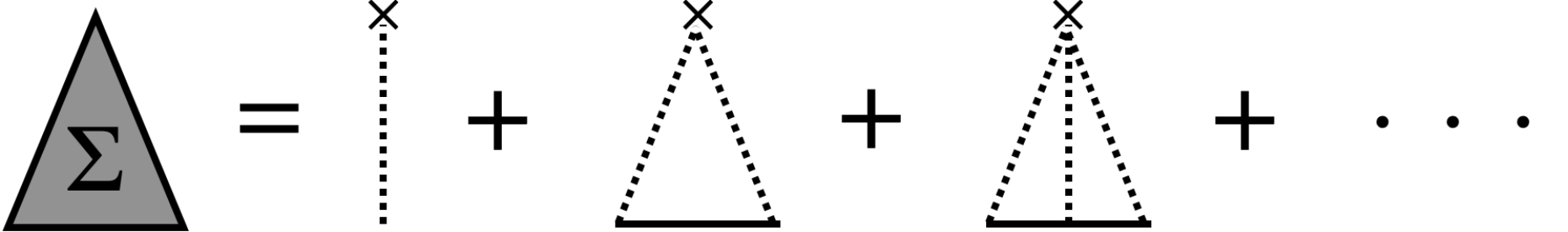}
  \caption{Feynman diagrams for the $T$-matrix approximation of the self-energy $\varSigma$. The cross represents an impurity. The solid line denotes the Green's function, and the dotted line represents the impurity scattering potential.}\label{fig:self-energy}
\end{figure}

A straightforward approach to incorporating the field dependence of the scattering rate is to calculate the self-energy using the Born approximation up to the second order of the interaction, $ \mathcal{O}(U^2) $, which is valid for relatively weak scatterings.
In fact, the constant relaxation-time approximation is effective at low magnetic fields, yielding good quantitative agreement with experiments~\cite{Collaudin2015-fk,Zhu_2018,Mitani_2020}. 
However, in the QL, the kinetic energy decreases rapidly as the magnetic field increases. 
As shown later, the Born approximation yields $1/\tau \propto B^2$ in this region.
Namely, the scattering rate is enhanced by the magnetic field, and the Born approximation would break down. 
Consequently, higher-order scattering processes must be considered in the QL.

In this study, we calculate the self-energy using the $T$-matrix approximation\cite{Altland_2010}, which systematically includes perturbation to infinite order for impurity scattering and is valid even for the strong scatterings, as illustrated in Fig.~\ref{fig:self-energy}.
The impurity scattering potential is assumed to be short-range: $\mathcal U(\bm r) = \sum_i U\delta(\bm r - \bm r_i)$, where $U$ is the impurity potential strength and $\bm{r}_i$ is the position of the impurity. 
In this case, the self-energy within the $T$-matrix approximation satisfies
\begin{eqnarray}
        \varSigma(i\varepsilon_n) = n_i U - i\mathrm{sgn}(\varepsilon_n)U\frac{m^{\ast2}}{2\pi \hbar^4}\sum_{\ell}\frac{\hbar\omega_c }{k_\ell(i\varepsilon_n)} \varSigma(i\varepsilon_n),
\end{eqnarray}
where $k_\ell(i\varepsilon_n) = \sqrt{2m^\ast\left[i\varepsilon_n - \left( \ell + \frac{1}{2} \right)\hbar\omega_c\right]}/\hbar$ and $\varSigma^1 = n_iU$. 
In present calculations, the retarded self-energy at zero temperature, obtained by the analytic continuation $i\varepsilon_n \to \varepsilon + i\delta$ takes the form:
\begin{eqnarray}\label{eq:Abrikosov_self-energy}
  \varSigma^\mathrm{R}(\varepsilon) = \frac{n_i U}{1 + i U\frac{\displaystyle m^{\ast2}}{\displaystyle 2\pi\hbar^4}\sum_{\ell}\frac{\displaystyle \hbar\omega_c}{\displaystyle k_\ell(\varepsilon)}
  }.
\end{eqnarray}
Note that the second term of the denominator is proportional to the density of states. At zero temperature, only $\varSigma^{\rm R}(\mu)$ contributes to $\hbar /2\tau$ in Eqs. (\ref{eq:sigmaxx})-(\ref{eq:K_l}).

\begin{figure}[tbph]
  \centering
  \includegraphics[width=8.5cm]{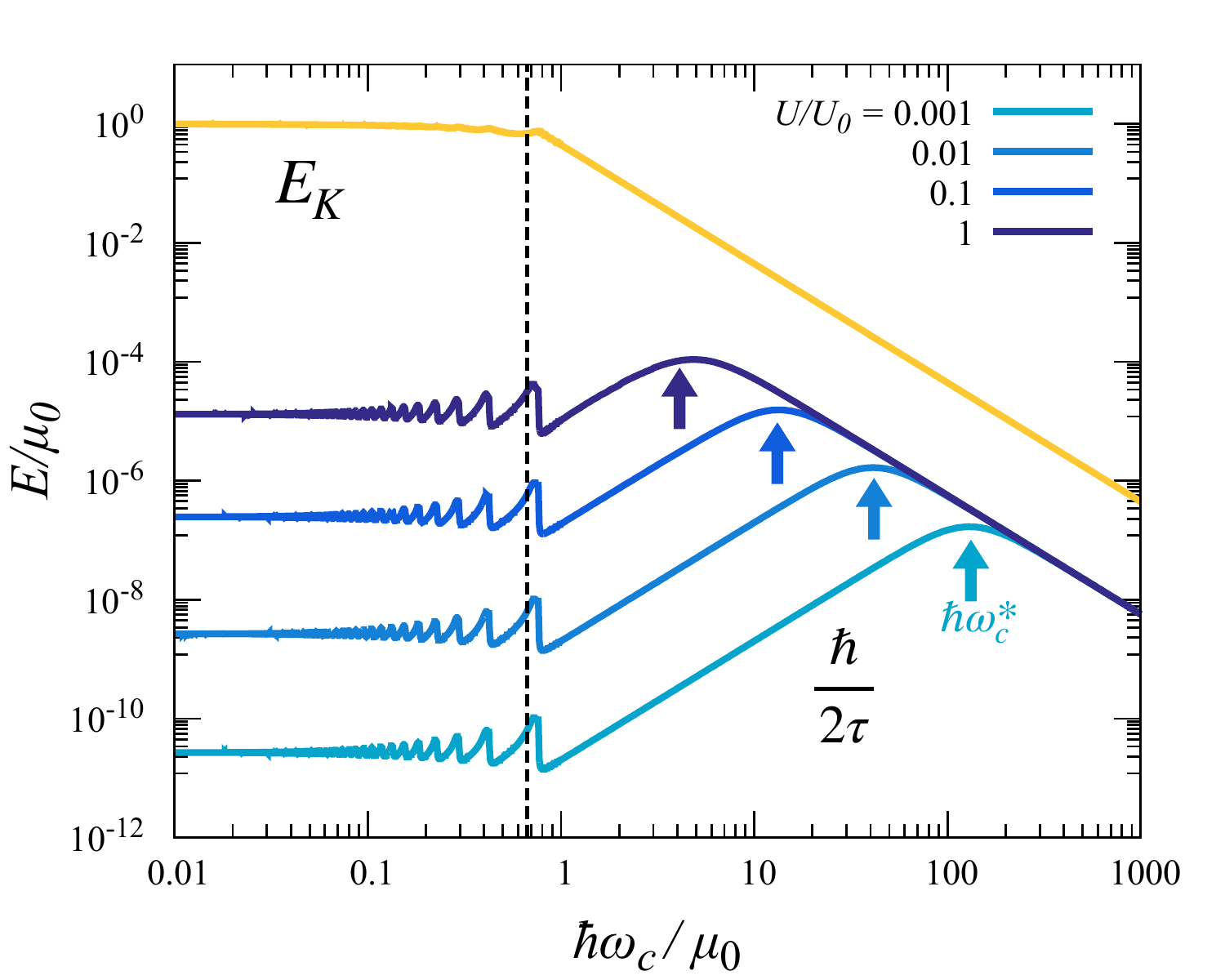}
  \caption{Magnetic field dependence of the kinetic energy $E_K$ and the scattering rate $\hbar /2\tau$ for different impurity scattering strengths $U$. Arrows indicate the crossover field $\hbar \omega_c^*$ for each value of $U$.
}\label{fig:kinetic_gamma}
\end{figure}

\section{Results and discussion}

\subsection{Kinetic energy and scattering rate}

The magnetic field dependence of the chemical potential and scattering rate is obtained by simultaneously solving the constant-carrier-density condition and the self-energy expression in Eq. (\ref{eq:Abrikosov_self-energy}) for various scattering potential intensities $U$, as shown in Fig.~\ref{fig:kinetic_gamma}.
Hereafter, we use the values $m^*=0.001 \,m_e$ for the effective mass; this is chosen because it corresponds to systems where the QL can be easily achieved by the current non-destructive magnetic field facilities. This is motivated by certain narrow-gap semiconductors and light-mass semimetals such as IV-VI semiconductors or bismuth~\cite{Zhu_2018}. $n_i = 0.01 \,n_0$ for the impurity density, and $\mu_0 = 30$ meV for the chemical potential. ($\hbar \omega_c /\mu_0 = 1.0$ corresponds to $B= 0.26$ T with these parameters.) The impurity potential strength is normalized by $U_0= \mu_0^{-1/2}(\hbar/\sqrt{2m^*})^3$.

The kinetic energy $E_K$  remains nearly constant below the QL, where $\hbar \omega_c < 2\mu_0/3$. However, beyond the QL, $E_K$ rapidly decreases as $E_K \propto B^{-2}$, to maintain a constant carrier density. This rapid decay of $E_K$ is a hallmark of the QL and plays a crucial role in producing anomalous magnetotransport phenomena, particularly in the EQL. 

The scattering rate $1/\tau$ also remains approximately constant below the QL, aside from quantum oscillations. This behavior strongly supports the validity of the constant-$\tau$ approximation in the sub-QL regime.
Beyond the QL, it initially increases as $1/\tau \propto B^2$, consistent with the Born approximation. 
However, at higher magnetic fields, $\hbar /2\tau$ never exceeds $E_K $. Instead, it turns to decrease, following the same field dependence as the kinetic energy: $1/\tau \propto B^{-2}$.
Notably, this behavior is independent of the scattering potential strength $U$; the scattering rate collapses onto a single universal line regardless of $U$.

\subsection{Unitarity limit and EQL}
The field dependence of the scattering rate discussed above reveals a qualitative transition in the scattering regime.
At lower fields in the QL, the system is well described by the Born approximation, where the scattering rate depends on $U$.
However, in the EQL, the scattering rate becomes independent of $U$, signaling a crossover to the unitarity limit.

$U$-independent scattering is a hallmark of the unitarity limit and is known to appear in various contexts such as superconductivity~\cite{Sigrist1991-sc,Lee1993-ie, Schmitt-Rink1986-dq}, the Kondo effect~\cite{Costi2000-dw}, two-dimensional graphene~\cite{Peres2006-qx, Kumazaki2006-ey}, and cold atoms~\cite{Giorgini2008-ji}, where the scattering rate is determined solely by the impurity type. 
In contrast, the present system exhibits a field-induced crossover from the Born regime to the unitarity limit --- driven purely by the magnetic field.

The origin of this crossover can be directly understood from Eq.~(\ref{eq:Abrikosov_self-energy}). When the second term in the denominator is small, the self-energy reduces to the Born form, and the scattering rate scales with $U$. In the opposite limit, where this term becomes large, the first term in the denominator becomes negligible, and the factor $U$ cancels between the numerator and the denominator. As a result, the scattering rate becomes inversely proportional to the density of states and independent of $U$, indicating the unitarity limit.

The crossover field $\hbar \omega_c^*$ at which this transition occurs can be estimated by requiring the second term in Eq.~(\ref{eq:Abrikosov_self-energy}) to be of order unity, yielding
\begin{eqnarray}
    \hbar\omega_c^* = \sqrt{\frac{2\pi}{3U}}\left( \frac{2\hbar^2\mu_0}{m^\ast} \right)^\frac{3}{4}.
\end{eqnarray}
To the best of our knowledge, this is the first demonstration of a magnetic-field-induced crossover from the Born regime to the unitarity limit.

The kinetic energy $E_K$ and the scattering rate $1/\tau$ in the EQL --- equivalently, in the unitarity limit --- can be analytically obtained by considering only the lowest Landau level $\ell = 0$. This approach provides accurate results in the EQL, yielding

\begin{eqnarray}
  E_K &= \mathrm{Re}\left[ K_0 \right] = \frac{4\pi^2\mu_0}{9\left[4\left( \frac{n_i}{n_0} \right)^2+\pi^2\right]} \left( \frac{\hbar\omega_c}{\mu_0} \right)^{-2} ,\label{eq:mu_EQL} \\ 
  \frac{\hbar}{2\tau} &= \frac{16\mu_0\left( \frac{n_i}{n_0} \right)}{9\sqrt{4\left( \frac{n_i}{n_0} \right)^2 + \pi^2}}\left( \frac{\hbar\omega_c}{\mu_0} \right)^{-2}. \label{eq:Gamma_EQL}
\end{eqnarray}
These analytic results clearly show that both $E_K$ and $1/\tau$ scale as $B^{-2}$, and that $1/\tau$ becomes independent of $U$ --- a definitive signature of the unitarity limit.
Furthermore, the ratio between the scattering rate and the kinetic energy is given by
\begin{eqnarray}
    \frac{\hbar /2\tau}{E_K }= \frac{4}{\pi}\frac{n_i}{n_0}\sqrt{1+\left(\frac{2}{\pi}\frac{n_i}{n_0} \right)^2}.
\end{eqnarray}
This ratio remains less than unity in the physically relevant regime $n_i \ll n_0$, indicating that scattering processes do not overwhelm the intrinsic kinetic energy of the electrons.
This condition is essential for maintaining coherence in electronic transport and confirms the physical consistency of the unitarity-limit picture in the EQL.

\subsection{Magnetoresistivity}

\begin{figure}[tb]
 \includegraphics[width=8cm] {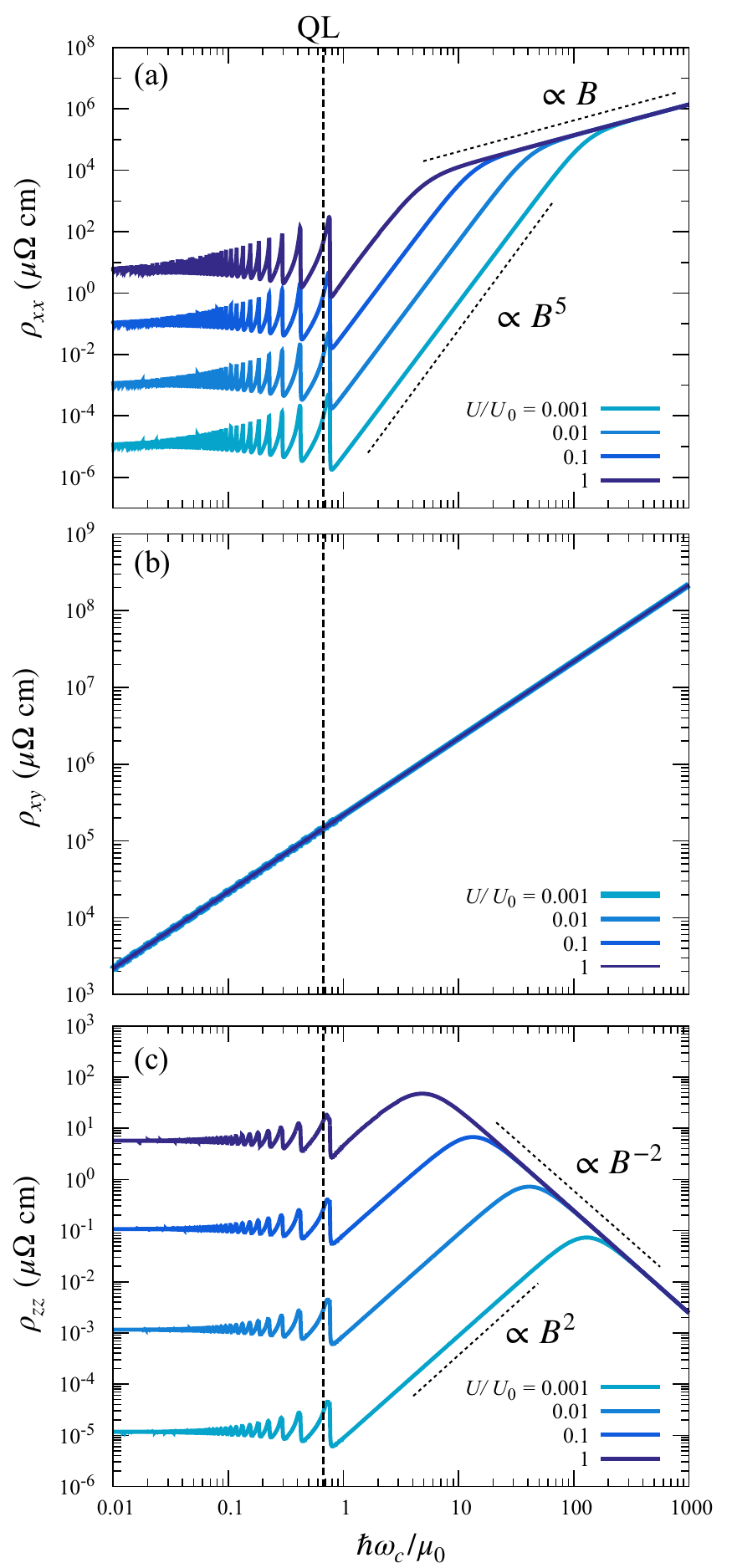}
  \caption{Magnetic field dependence of (a) the transverse magnetoresistivity $\rho_{xx}$, (b) the Hall resistivity $\rho_{xy}$, and (c) the longitudinal magnetoresistivity, for different impurity scattering strengths $U$. }\label{fig:rho}
\end{figure}

The transverse resistivity $\rho_{xx}$ and Hall resistivity $\rho_{xy}$ are shown in Fig. \ref{fig:rho}, where the magnetic field is applied parallel to the $z$-axis.
Below the QL, $\rho_{xx}$ remains approximately constant, aside from the quantum oscillation, in agreement with classical theory, namely, the quantum--classical correspondence holds~\cite{Yamada_2024}.
In the QL, however, $\rho_{xx}$ exhibits a clear crossover associated with the crossover from the Born regime to the unitarity limit.
In the Born regime of QL ($B<B^* \equiv m^* \omega_c^*/e$ ) $\rho_{xx}$ increases rapidly as $\rho_{xx}\propto B^5$.
In contrast, beyond $B^*$, the system enters the unitarity limit, where $\rho_{xx}$ transitions to a linear dependence, $\rho_{xx} \propto B$, which is a hallmark of the EQL. This linear transverse magnetoresistivity in the EQL agrees with Abrikosov's prediction under the condition $n_i \ll n_0$~\cite{Abrikosov_1969}, which was previously derived only in the high field limit. Our theory extends this result, providing a unified and quantitative description of magnetoresistivity across the entire field range, from weak fields to the EQL. As a result, the crossover behavior of $\rho_{xx}$ is clearly captured, revealing the field-induced crossover between fundamentally different scattering regimes.

In the EQL, $\rho_{xx}$ collapses onto a single universal line, independent of the impurity potential strength, which is also the hallmark of the unitarity limit or the EQL.
The analytic form of $\rho_{xx}$ in the EQL is given by
\begin{eqnarray}
  \rho_{xx}^{\mathrm{EQL}} &=& \frac{2B}{en_0}\frac{\left( \frac{n_i}{n_0} \right)}{\sqrt{4\left( \frac{n_i}{n_0} \right)^2 + \pi^2}}.\label{eq:rho_xx}
\end{eqnarray}
It is clear from the above formula that $\rho_{xx}$ is independent of the scattering strength $U$, and it depends only on $n_0$, $n_i$, and $B$. Consequently, it enables experimental determination of $n_i$ through direct measurement of $\rho_{xx}$ and $\rho_{xy}$, as will be discussed later.

In contrast, the Hall resistivity $\rho_{xy}$ exhibits neither the quantum oscillations below the QL nor any crossover behavior in the QL. Instead, it maintains the universal relation:
\begin{eqnarray}
  \rho_{xy} &=& \frac{B}{en_0}, \label{eq:rho_xy}
\end{eqnarray}
which holds across the entire field range.

The longitudinal resistivity, $\rho_{zz}$, exhibits distinct behavior.
Below the QL, it remains nearly constant except for quantum oscillations [Fig. \ref{fig:rho}(c)]. 
In the QL, $\rho_{zz}$ also undergoes a crossover: $\rho_{zz}$ increases as $\rho_{zz}\propto B^2$ in the Born regime, while it decreases as $\rho_{zz}\propto B^{-2}$ in the unitarity limit ($B>B^*$). This field-dependent crossover leads to negative longitudinal magnetoresistance in the EQL --- a clear signature of the crossover between scattering regimes. 
So far, nothing unusual has been expected in the QL in the absence of electron-electron correlations. However, our results indicate a magnetic-field-induced crossover from the Born regime to the unitarity limit, as revealed by the behavior of $\rho_{xx}$ and $\rho_{zz}$.

Thus, the transverse and longitudinal magnetoresistivities exhibit strikingly different behaviors in the EQL:
\begin{itemize}
    \item Transverse resistivity $\rho_{xx}$ increases linearly with $B$.
    \item Longitudinal resistivity $\rho_{zz}$ decreases with $B^{-2}$.
\end{itemize}
This anisotropy implies that the system behaves as an insulator when the field is perpendicular to the current but as a metal when the field is parallel to the current.
Consequently, the resistivity can be drastically modified by rotating the magnetic field direction --- an essential characteristic of the EQL. This is the answer to the question raised in the Introduction.

\subsection{Measurement for impurity density}

\begin{figure}[tb]
  \centering
  \includegraphics[width = 8cm]{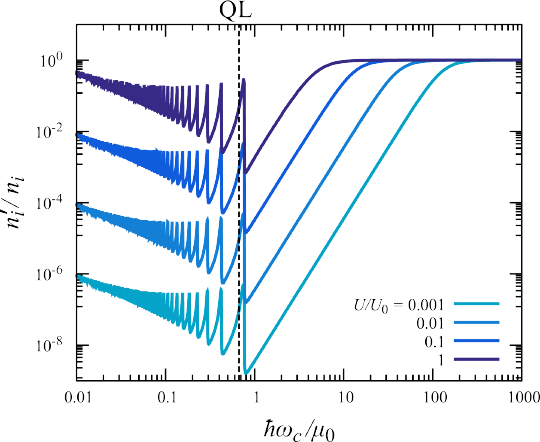}
  \caption{Magnetic field dependence of $n_i'=(\pi B/2e)(\rho_{xx}/\rho_{xy})\left( \rho_{xy}^2-\rho_{xx}^2\right)^{-1/2}$ for different impurity scattering strengths $U$. $n_i'$ matches with the true impurity density $n_i$ in the EQL (or in the unitarity limit).}\label{fig:impurity-density}
\end{figure}

In the unitarity limit, the energy scale associated with the scattering potential $U$ no longer plays a role. This allows us to derive a universal expression that is independent of the microscopic details of the impurity scattering.
By combining Eqs.~(\ref{eq:rho_xx}) and (\ref{eq:rho_xy}), we obtain a new formula for evaluating the impurity density in the EQL:
\begin{eqnarray}
  n_i = \lim_{B\to {\rm EQL}}\frac{\pi B}{2e}\frac{\rho_{xx}}{\rho_{xy}}\left( \rho_{xy}^2 - \rho_{xx}^2 \right)^{-1/2}.\label{eq:impurity}
\end{eqnarray}
This result demonstrates that the impurity density can be quantitatively determined by simultaneous measurements of $\rho_{xx}$ and $\rho_{xy}$ in the EQL. This approach is analogous to the standard evaluation of carrier density via Hall coefficient measurements. However, it provides a direct measurement of impurity density, which is typically challenging to evaluate using conventional techniques. This is a direct consequence --- and a notable advantage --- of reaching the unitarity limit.
Figure \ref{fig:impurity-density} shows the field dependence of $n_i'=(\pi B/2e)(\rho_{xx}/\rho_{xy})\left( \rho_{xy}^2-\rho_{xx}^2\right)^{-1/2}$.
In the EQL, $n_i'$ becomes independent of $B$ and converges to the actual impurity $n_i$ in the system.
This universality of $n_i$ in the EQL highlights a practical advantage of our theory: the impurity density can be experimentally determined.

\subsection{Quantum--classical correspondence}
The field dependence of magnetoresistance (MR) can be naturally understood through the quantum--classical correspondence.
For clean systems $\hbar/2\tau \ll E_K$, Eqs. (\ref{eq:sigmaxx})-(\ref{eq:sigmazz}) can be rewritten as
\begin{eqnarray}
  \sigma_{xx} &=& \frac{n_0 e^2\tau}{m}\frac{Q(B)}{(\omega_c\tau)^2} 
  ,\label{eq:sigmaxx_2} \\
  \sigma_{yx}&=& \frac{en_0}{B}
  ,\label{eq:sigmayx_2} \\
  \sigma_{zz} &=& \frac{n_0 e^2\tau}{m},\label{eq:sigmazz_2}
\end{eqnarray}
where
\begin{eqnarray}
  Q(B) = \frac{3\pi\hbar^2}{2m^*\mu_0^2}\sum_\ell N_L \left(\ell + \frac{1}{2}\right)\hbar\omega_c\mathrm{Re}\left[\sqrt\frac{\mu_0}{{K_\ell}}\right]\label{eq:Q}.
\end{eqnarray}
The ${\rm Im} [\cdots]$ terms in Eqs. (\ref{eq:sigmayx}) and (\ref{eq:sigmazz}) are negligible in this regime and are therefore discarded. 
Remarkably, the above expressions agree with the classical magnetoconductivity formulas, except for the quantum correction factor $Q(B)$. 

\begin{figure}[tb]
  \centering
  \includegraphics[width = 8cm]{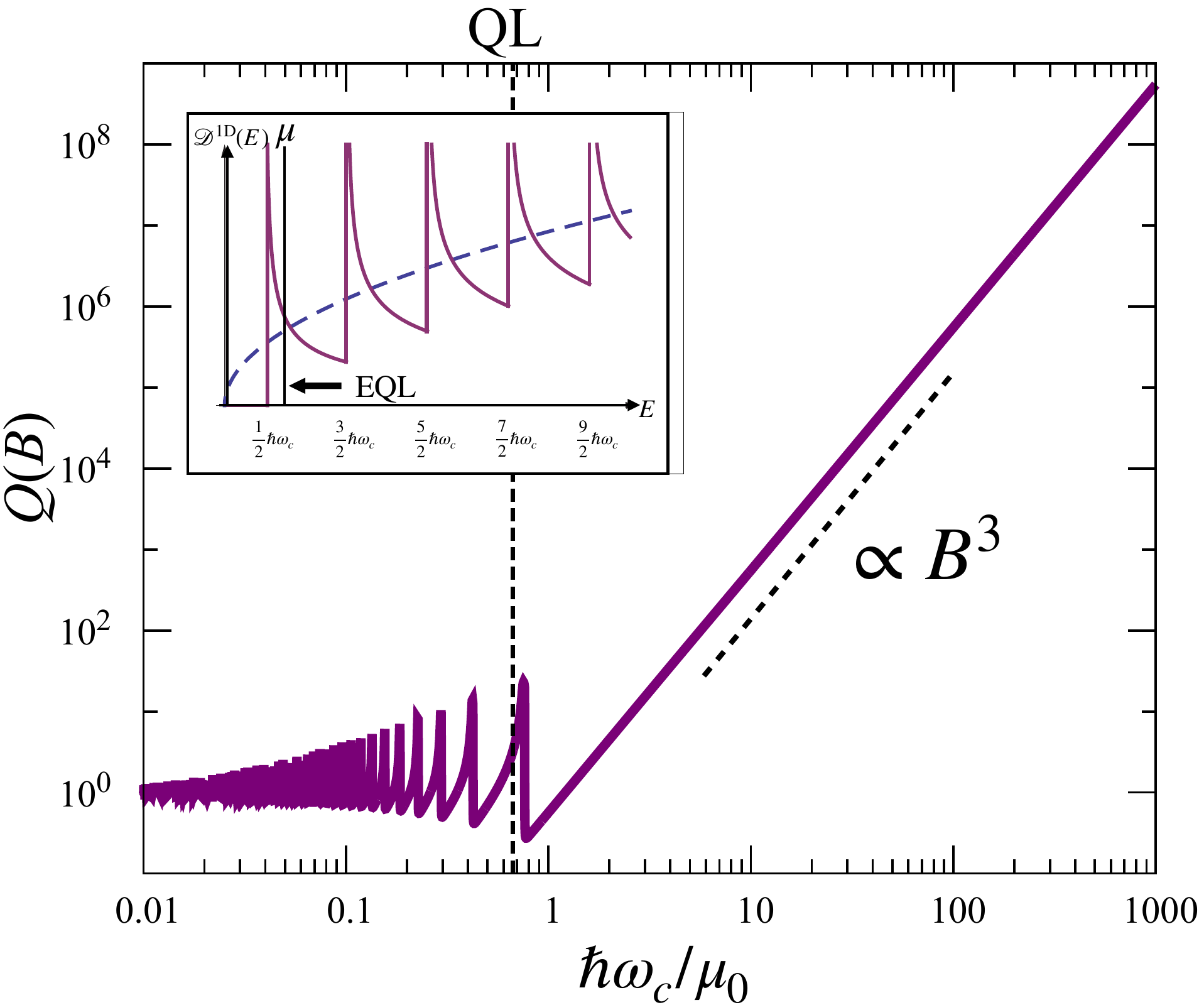}
  \caption{Magnetic field dependence of the quantum correction factor $Q(B)$. The inset shows the density of states under a strong magnetic field, clearly exhibiting Landau quantization.}\label{fig:Q}
\end{figure}

Below the QL, $Q(B)$ merely contributes to quantum oscillations, consistent with previous results~\cite{Yamada_2024}.
However, in the QL, $Q(B)$ exhibits a dramatic change: it scales as $Q(B) \propto B^3$ as shown in Fig. \ref{fig:Q}.
The quantum correction $Q(B)$ consists of three factors:
\begin{eqnarray}
    Q(B) &\propto \hbar \omega_c \cdot N_L \cdot D_{\rm 1D}
    \\
    &\propto B\cdot B\cdot B \propto B^3, \nonumber
\end{eqnarray}
where the cyclotron energy scales linearly as $\hbar \omega_c \propto B$, the Landau degeneracy increases linearly as $N_L \propto B$, and the one-dimensional density of states scales as $D_{\rm 1D}\propto E_K^{-1/2}\propto B$ due to the relationship $E_K\propto B^{-2}$ from Eq. (\ref{eq:mu_EQL}).
Thus, $Q(B) \propto B^3$.
This scaling of $Q(B)$ plays a crucial role in determining the magnetoresistive behavior in the EQL.

The quantum--classical correspondence allows us to understand the field dependence of magnetoconductivities by considering the field dependence for the scattering rate and quantum correction:
\begin{eqnarray}
    \frac{1}{\tau} \propto B^{-2}, \quad Q(B)\propto B^3.
\end{eqnarray}
Applying these dependences to Eqs.~(\ref{eq:sigmaxx_2})-(\ref{eq:sigmazz_2}) yields:
\begin{eqnarray}
  \sigma_{xx} &= \frac{n_0e^2}{m}\frac{Q(B)}{\omega_c^2 \tau} \propto B^{-1}, \\
  \sigma_{yx} &=\frac{en_0}{B} \propto B^{-1}, \\
  \sigma_{zz} &= \frac{n_0 e^2 \tau}{m} \propto B^2. \label{eq:sigmazz_3}
\end{eqnarray}
These results provide precise explanations for the observed magnetoresistivities in the EQL: $\rho_{xx} \propto B$, $\rho_{xy}\propto B$, and $\rho_{zz}\propto B^{-2}$. 

The nature of the negative longitudinal MR $\rho_{zz}\propto B^{-2}$ can be understood as follows.
The longitudinal MR is proportional to the scattering rate as $\rho_{zz} \propto 1/\tau$ [cf. Eq. (\ref{eq:sigmazz_2})].
In the EQL, the scattering rate is inversely proportional to the density of state ($ = N_L \cdot D_{\mathrm{1D}}$), which increases in the QL. Namely, the scattering rate decreases with the magnetic field as $1/\tau \propto B^{-2}$, yielding the decrease of longitudinal MR as $\rho_{zz} \propto B^{-2}$.

The quantum--classical correspondence provides a unified framework to understand magnetoresistivity even in the EQL. By appropriately accounting for the field dependences of $\tau$ and $Q(B)$, we have shown that classical-like expressions for conductivity remain valid with quantum corrections.

\subsection{Realistic systems}

In realistic systems, the Coulomb interaction may induce some type of collective transition in a strong magnetic field~\cite{Halperin1987-xd}. For example, in graphite, Coulomb interactions can induce charge density wave (CDW) order or excitonic states in the extreme quantum limit\cite{Zhu_undated-sj, Akiba2015-lk, Arnold2017-um}. 
In contrast, in systems with high dielectric constants, Coulomb interactions are strongly screened, which is believed to suppress electron correlation effects. Bismuth is a representative example of such a system, and no Coulomb interaction-induced phenomena have been reported, even under extremely strong magnetic fields ($\sim 90$ T) \cite{Zhu2017-bb}. 
The present theory is applicable to such systems where the impurity scattering is dominant, and the effect of the Coulomb interaction is suppressed due to the screening.

Negative longitudinal MR and linear transverse MR have been observed in various materials \cite{Wiedmann2016-qz, Li2016-zp, Zhang2017-lu, Hu2008-jk, Novak2015-bo, Narayanan2015-bo, Zhu2017-bb, Balduini2024-qw}, including topological systems. It remains necessary to carefully examine whether these phenomena originate from the present mechanism in the EQL. Importantly, the negative longitudinal MR reported here emerges only in the strong-field regime where the system enters the unitarity limit. In contrast, negative MR associated with the chiral anomaly can occur even at weak magnetic fields~\cite{Li2016-zp, Wiedmann2016-qz, Balduini2024-qw}. These distinct field regimes allow clear differentiation between the two mechanisms.

Systems that can realistically reach the EQL are limited to those with a small carrier density and a low effective mass. Bismuth is a representative example, and is known to the Dirac electron systems with a light effective mass. The effective mass of Dirac electrons is typically defined within the non-relativistic approximation \cite{Wolff1964-wc}. In this framework, the effective mass corresponds to the curvature of the band edge and remains constant. Moreover, the interband effects are expected not to play a significant role when the kinetic energy is sufficiently less than the band gap, at least for the magnetoresistance in the Dirac electron systems \cite{Owada2018-jd}.
Consequently, the present theory is applicable to Dirac electron systems in the EQL. 
Linear MR has been observed in bismuth under strong magnetic fields above 60 T. The mechanism of this linear MR can originate from the characteristics of the EQL discussed in this study.
However, the field dependence of chemical potential in Bi is very complex due to the semimetallic nature.
Furthermore, although Eq.~(\ref{eq:impurity}) is only valid when the Hall resistivity is larger than the transverse magnetoresistivity, the Hall resistivity is generally small due to the presence of multiple carrier types in a semimetal.
Therefore, an extension of the present theory to semimetals would allow quantitative comparison with experiments.

Another important extension of our analysis involves considering a magnetic field that is not perpendicular or parallel to the electric current direction. 
When the magnetic field is rotated from being perpendicular to parallel to the electric current, the system will exhibit a crossover in MR, i.e., crossover from insulating behavior characterized by linear transverse MR to metallic behavior characterized by negative longitudinal MR. This anisotropic response is highly nontrivial and can be directly verified through experiments.

\section{Conclusion}
In this study, we have theoretically investigated the magnetoresistance (MR) in the extreme quantum limit (EQL) by deriving transverse, Hall, and longitudinal resistivities ($\rho_{xx}$, $\rho_{xy}$, and $\rho_{zz}$) using the Kubo formula with Green's functions.
The scattering rate $1/\tau$ has been calculated within the $T$-matrix approximation, enabling a consistent description of transport from weak magnetic fields up to the EQL.

A key finding is the emergence of a magnetic-field-induced crossover in the scattering rate within the quantum limit (QL), a regime where no anomaly is typically expected according to conventional theories.
In the Born regime ($B<B^*$), the scattering rate increases as $1/\tau \propto B^2$, while in the unitarity limit ($B>B^*$), it decreases as $1/\tau \propto B^{-2}$.
This identifies the EQL as a realization of the unitarity limit, where scattering reaches the maximum allowed by quantum mechanics~\cite{Sakurai2020-hd}, and transport becomes governed by universal relations that are independent of the microscopic details of impurity scattering.

This crossover in $1/\tau $ directly manifests in the transverse MR $\rho_{xx}$ and the longitudinal MR $\rho_{zz}$.
In the Born regime, $\rho_{xx} \propto B^5$ and $\rho_{zz} \propto B^2$, whereas in the unitarity limit, $\rho_{xx} \propto B$ and $\rho_{zz} \propto B^{-2}$.
Consequently, the EQL is characterized by a linear transverse MR and a negative longitudinal MR.
This distinct anisotropy implies that the system behaves as an insulator when the field is perpendicular to the current, but as a metal when the field is parallel to the current. Such a significant anisotropy is a unique feature of the EQL.
We have also confirmed that the quantum--classical correspondence remains valid even in the EQL when the field dependencies of $\tau$ and $Q(B)$ are appropriately taken into account.

Furthermore, we have derived a new universal formula, Eq.~(\ref{eq:impurity}), which enables the experimental determination of the impurity density by simultaneously measuring $\rho_{xx}$ and $\rho_{xy}$ in the EQL.
This approach provides a practical and quantitative method for probing impurity concentrations in systems where conventional methods are often ineffective.

Importantly, the field-induced crossover to the unitarity limit is expected to be a general feature, independent of the specific theoretical treatment of impurity scattering.
The Born approximation yields accurate results in the weak scattering regime; however, it eventually breaks down at high magnetic fields, where the effective scattering strength is amplified. However, the enhancement of the scattering amplitude is fundamentally constrained by quantum mechanics. As a result, the unitarity limit emerges as a universal fixed point in the EQL, irrespective of the details of the approximation. While the specific $B$-dependence of MR may vary across models, the existence of the crossover should be a robust and model-independent feature.

Nevertheless, the presence of field-induced phase transitions driven by electron-electron interactions, neglected in the present analysis, could potentially mask the crossover behavior~\cite{Halperin1987-xd}. The relevance of such competing effects depends on the material; for instance, graphite undergoes a transition under high fields~\cite{Zhu_undated-sj, Akiba2015-lk, Arnold2017-um}, whereas bismuth does not ~\cite{Zhu2017-bb}. In any case, the exploration of the EQL remains in its infancy, offering fertile ground for discovering new quantum phenomena.

\section*{Acknowledgments}
We thank M. Tokunaga, Y. Matsuda, H. Sakai, and T. Yamaguchi for the fruitful discussions. This work was supported by the Japan Society for the Promotion of Science (Grants No. 23H00268, No. 23H04862, and No. 22K18318).

\section*{References}
\bibliographystyle{iopart-num}
\bibliography{eql}

\providecommand{\newblock}{}
\begin{thebibliography}{10}
\expandafter\ifx\csname url\endcsname\relax
  \def\url#1{{\tt #1}}\fi
\expandafter\ifx\csname urlprefix\endcsname\relax\def\urlprefix{URL }\fi
\providecommand{\eprint}[2][]{\url{#2}}

\bibitem{Landau1930}
Landau L 1930 {\em Zeitschrift f{\"u}r Physik\/} {\bf 64} 629--637

\bibitem{Shoenberg_book}
Shoenberg D 1984 {\em Magnetic Oscillations in Metals\/} Cambridge Monographs on Physics (Cambridge University Press)

\bibitem{Hu2008-jk}
Hu J and Rosenbaum T~F 2008 {\em Nat. Mater.\/} {\bf 7} 697--700

\bibitem{Bhattacharya2016-yv}
Bhattacharya A, Skinner B, Khalsa G and Suslov A~V 2016 {\em Nat. Commun.\/} {\bf 7} 12974

\bibitem{Kozii2019-ca}
Kozii V, Skinner B and Fu L 2019 {\em Phys. Rev. B.\/} {\bf 99} 155123

\bibitem{Konye_2018}
Könye V and Ogata M 2018 {\em Phys. Rev. B.\/} {\bf 98} 195420

\bibitem{Kubo_formula}
Kubo R 1957 {\em J. Phys. Soc. Jpn.\/} {\bf 12} 570--586

\bibitem{Kubo_solid}
Kubo R, Miyake S~J and Hashitsume N 1965 Quantum theory of galvanomagnetic effect at extremely strong magnetic fields ({\em Solid state physics\/} vol~17) ed Seitz F and Turnbull D (Elsevier) pp 269--364

\bibitem{Fukuyama_1969}
Fukuyama H, Ebisawa H and Wada Y 1969 {\em Prog. Theor. Phys.\/} {\bf 42} 494--511

\bibitem{Shiba_1971}
Shiba H, Kanda K, Hasegawa H and Fukuyama H 1971 {\em J. Phys. Soc. Jpn.\/} {\bf 30} 972--987

\bibitem{Fuseya_2015}
Fuseya Y, Ogata M and Fukuyama H 2015 {\em J. Phys. Soc. Jpn.\/} {\bf 84} 012001

\bibitem{Yamada_2024}
Yamada A and Fuseya Y 2024 {\em J. Phys.: Condens. Matter\/} {\bf 36} 245702

\bibitem{Luttinger1955-nd}
Luttinger J~M and Kohn W 1955 {\em Phys. Rev.\/} {\bf 97} 869--883

\bibitem{Wolff1964-wc}
Wolff P~A 1964 {\em J. Phys. Chem. Solids\/} {\bf 25} 1057--1068

\bibitem{Matsubara_1955}
Matsubara T 1955 {\em Prog Theor Phys\/} {\bf 14} 351--378

\bibitem{AGD}
Abrikosov A~A, Gorkov L~P and Dzyaloshinski I~E 1975 {\em Methods of Quantum Field Theory in Statistical Physics\/} (Dover Publications)

\bibitem{mahan}
Mahan G 2014 {\em Many-Particle Physics\/} Physics of Solids and Liquids (Springer)

\bibitem{rickayzen_1980}
Rickayzen G 1980 {\em Green's Functions and Condensed Matter\/} Techniques of Physics (Academic Press)

\bibitem{Collaudin2015-fk}
Collaudin A, Fauqué B, Fuseya Y, Kang W and Behnia K 2015 {\em Phys. Rev. X.\/} {\bf 5} 021022

\bibitem{Zhu_2018}
Zhu Z, Fauqu{\'{e}} B, Behnia K and Fuseya Y 2018 {\em J. Phys.: Condens. Matter\/} {\bf 30} 313001

\bibitem{Mitani_2020}
Mitani Y and Fuseya Y 2020 {\em J. Phys.: Condens. Matter\/} {\bf 32} 345802

\bibitem{Altland_2010}
Altland A and Simons B~D 2010 {\em Condensed matter field theory\/} 2nd ed (Cambridge, England: Cambridge University Press)

\bibitem{Sigrist1991-sc}
Sigrist M and Ueda K 1991 {\em Rev. Mod. Phys.\/} {\bf 63} 239

\bibitem{Lee1993-ie}
Lee P~A 1993 {\em Phys. Rev. Lett.\/} {\bf 71} 1887--1890

\bibitem{Schmitt-Rink1986-dq}
Schmitt-Rink S, Miyake K and Varma C~M 1986 {\em Phys. Rev. Lett.\/} {\bf 57} 2575--2578

\bibitem{Costi2000-dw}
Costi T~A 2000 {\em Phys. Rev. Lett.\/} {\bf 85} 1504--1507

\bibitem{Peres2006-qx}
Peres N~M~R, Castro~Neto A~H and Guinea F 2006 {\em Phys. Rev. B\/} {\bf 73} 195411

\bibitem{Kumazaki2006-ey}
Kumazaki H and S~Hirashima D 2006 {\em J. Phys. Soc. Jpn.\/} {\bf 75} 053707

\bibitem{Giorgini2008-ji}
Giorgini S, Pitaevskii L~P and Stringari S 2008 {\em Rev. Mod. Phys.\/} {\bf 80} 1215--1274

\bibitem{Abrikosov_1969}
Abrikosov A~A 1969 {\em Sov. Phys. JETP\/} {\bf 29} 746

\bibitem{Halperin1987-xd}
Halperin B~I 1987 {\em Jpn. J. Appl. Phys.\/} {\bf 26} 1913

\bibitem{Zhu_undated-sj}
Zhu Z, Nie P, Fauqué B, Vignolle B, Proust C, McDonald R~D, Harrison N and Behnia K 2019 {\em Physical Review X\/} {\bf 9} 011058

\bibitem{Akiba2015-lk}
Akiba K, Miyake A, Yaguchi H, Matsuo A, Kindo K and Tokunaga M 2015 {\em J. Phys. Soc. Jpn.\/} {\bf 84} 054709

\bibitem{Arnold2017-um}
Arnold F, Isidori A, Kampert E, Yager B, Eschrig M and Saunders J 2017 {\em Phys. Rev. Lett.\/} {\bf 119} 136601

\bibitem{Zhu2017-bb}
Zhu Z, Wang J, Zuo H, Fauqué B, McDonald R~D, Fuseya Y and Behnia K 2017 {\em Nat. Commun.\/} {\bf 8} 15297

\bibitem{Wiedmann2016-qz}
Wiedmann S, Jost A, Fauqué B, van Dijk J, Meijer M~J, Khouri T, Pezzini S, Grauer S, Schreyeck S, Brüne C, Buhmann H, Molenkamp L~W and Hussey N~E 2016 {\em Phys. Rev. B.\/} {\bf 94} 081302

\bibitem{Li2016-zp}
Li H, He H, Lu H~Z, Zhang H, Liu H, Ma R, Fan Z, Shen S~Q and Wang J 2016 {\em Nat. Commun.\/} {\bf 7} 10301

\bibitem{Zhang2017-lu}
Zhang C~L, Schindler F, Liu H, Chang T~R, Xu S~Y, Chang G, Hua W, Jiang H, Yuan Z, Sun J, Jeng H~T, Lu H~Z, Lin H, Hasan M~Z, Xie X~C, Neupert T and Jia S 2017 {\em Phys. Rev. B\/} {\bf 96} 165148

\bibitem{Novak2015-bo}
Novak M, Sasaki S, Segawa K and Ando Y 2015 {\em Phys. Rev. B\/} {\bf 91} 041203

\bibitem{Narayanan2015-bo}
Narayanan A, Watson M~D, Blake S~F, Bruyant N, Drigo L, Chen Y~L, Prabhakaran D, Yan B, Felser C, Kong T, Canfield P~C and Coldea A~I 2015 {\em Phys. Rev. Lett.\/} {\bf 114} 117201

\bibitem{Balduini2024-qw}
Balduini F, Molinari A, Rocchino L, Hasse V, Felser C, Sousa M, Zota C, Schmid H, Grushin A~G and Gotsmann B 2024 {\em Nat. Commun.\/} {\bf 15} 6526

\bibitem{Owada2018-jd}
Owada M, Awashima Y and Fuseya Y 2018 {\em J. Phys. Condens. Matter\/} {\bf 30} 445601

\bibitem{Sakurai2020-hd}
Sakurai J~J and Napolitano J 2020 {\em Modern quantum mechanics\/} 3rd ed (Cambridge, England: Cambridge University Press)

\end{thebibliography}

\end{document}